\documentclass[12pt,preprint]{aastex}

\newcommand{\etal}{{et~al.~}}

\newcommand{\co}{\mbox{$^{12}$CO}}
\newcommand{\coa}{\mbox{$^{13}$CO~}}

\newcommand{\cmsq}{\mbox{${\rm cm}^{-2}$}}
\newcommand{\kms}{\mbox{${\rm km~s}^{-1}$}}

\newcommand{\vlsr}{\mbox{$\rm V_{\rm LSR}$~}}

\newcommand{\htwo}{\mbox{${\rm H}_2$~}}

\newcommand{\msun}{\mbox{$M_\odot$~}}

\shorttitle{Turbulence in the Rosette and G216-2.5  Molecular Clouds}
\shortauthors{Heyer et al.}

\begin{document}


\title{Turbulent Gas Flows in the Rosette and G216-2.5 Molecular Clouds:
Assessing Turbulent Fragmentation Descriptions of Star Formation}

\author{Mark H. Heyer\altaffilmark{1}, Jonathan P. Williams\altaffilmark{2}, 
and
Christopher M. Brunt\altaffilmark{1,3}}


\begin{abstract}
The role of turbulent fragmentation in regulating the efficiency
of star formation in interstellar clouds is examined from new 
wide field imaging of \co\ and \coa\ J=1-0 emission from the Rosette
and G216-2.5 molecular clouds.  The Rosette molecular cloud is a typical
star forming giant molecular cloud and G215-2.5 is a massive molecular 
cloud with no OB stars and very little low mass star formation.
The properties
of the turbulent gas flow are derived 
from the set of eigenvectors and eigenimages
generated by Principal Component Analysis of the spectroscopic data cubes.
While the two clouds represent quite divergent states of 
star formation activity,
the velocity structure functions for both clouds are similar.  The sonic scale,
$\lambda_S$, 
defined as the spatial scale at which turbulent velocity fluctuations 
are equivalent to the local sound speed,
and the turbulent Mach number evaluated at 1 pc, $M_{1pc}$,
are derived for an ensemble 
of clouds including the Rosette and, 
G216-2.5 regions that span a large range in star formation activity.
We find no evidence for the positive correlations 
between these quantities and the star formation efficiency,
that are predicted by turbulent fragmentation models.
A correlation does exist between the star formation efficiency and the sonic scale 
for a subset of clouds with 
$L_{FIR}/M(H_2) > 1$ that are generating young stellar clusters.
Turbulent fragmentation must play a limited and 
non-exclusive role in determining the yield of 
stellar masses within interstellar clouds. 
\end{abstract}
\keywords{ISM: Individual: (Rosette Molecular Complex, G216-2.5) 
-- ISM:hydrodynamics ---
ISM: kinematics and dynamics --- ISM: clouds
stars: formation -- turbulence
}

\altaffiltext{1}{Department of Astronomy, University of Massachusetts,
    Amherst, MA 01002, {heyer@astro.umass.edu} }
\altaffiltext{2}{Institute for Astronomy, University of Hawaii, Honolulu, HI, 96822,
 {jpw@ifa.hawaii.edu}   }
\altaffiltext{3}{School of Physics, University of Exeter, Stocker Road,
EX4 4QL, United Kingdom, {brunt@astro.ex.ac.uk}}

\section{Introduction}

Newborn stars emerge from the cold, dense, molecular phase of the 
interstellar medium (ISM).  The efficiency through which stars form 
and the mass distribution of the stellar product have important 
implications to the galactic environment and evolution. 
The global star formation efficiency, defined as the ratio of mass of 
newborn stars and the mass of the parent molecular cloud, is observed to be 
low (several percent) 
and provides a fundamental observational 
constraint to descriptions of star formation and interstellar gas dynamics.

The inefficiency of star formation has been attributed to 
the quasi-static 
equilibrium of the molecular gas 
in which self-gravity is balanced by magnetic tension (Mouschovias 1976; Shu, Li, \& Allen 2004; 
Nakamura \& Li 2005).  
In this state, star formation is limited to those regions within a cloud in which 
the mass to flux ratio, $M/\Phi$, exceeds the critical value, 
$(M/\Phi)_{crit}=(4{\pi}^2G)^{-1/2}$
(Nakano \& Nakamura 1978).  Ambipolar diffusion provides a mechanism 
to increase $M/\Phi$ beyond the critical value within a region enabling the 
production of stars 
(Lizano \& Shu 1989).  

More 
recent descriptions of star formation 
have proposed that density enhancements from which new stars condense
within molecular clouds are generated by shocks driven by supersonic turbulent 
flows (Padoan \& Nordlund 2002; 
Mac Low \& Klessen 2004).  Whether such shock-induced 
compressions lead to self-gravitating protostellar cores that are dynamically 
decoupled from the overlying turbulent flow 
depends on specific 
properties of the velocity field. Klessen, Heitsch, \& Mac Low (2000) 
demonstrate a relationship between the star formation efficiency and the scale,
$\lambda_D$, at which 
the turbulent velocity field is driven.  The driving scale sets the time 
interval  between successive shocks.  If sufficiently long, the 
initial compression has time to evolve into a higher density, self-gravitating 
configuration that is stable against a subsequent shock and can 
more likely 
develop into a protostellar core. 
The sonic scale, $\lambda_S$, is defined as the scale at which the turbulent 
velocity fluctuations are equal to the sound speed and sets the spatial 
regime at which shocks can occur ($L > \lambda_S$).  
Vazquez-Semadeni, Ballesteros-Paredes, \& Klessen (2003) predict more efficient 
star formation with increasing sonic scale as there is a larger reservoir of 
material within a thermal core that is susceptible to local collapse (Padoan 1995).
For an isothermal gas, the turbulent Mach number, $M_s$, sets the amplitude of the shock.
Deeper compressions from larger Mach number turbulence 
can more rapidly evolve into dense, self-gravitating cores that are more likely to 
form stars.  One would expect a positive correlation of star formation 
efficiency with Mach number (Mac Low \& Klessen 2004; Nakamura \& Li 2005).  
For MHD turbulence in which 
the shocks are softened by the magnetic field, the thermal Mach number should be 
replaced with the Alfvenic Mach number. 

This dynamical view of star formation, 
labeled as ``turbulent fragmentation'' by 
its proponents, offers a 
compelling description of star formation but one that requires further scrutiny by 
observations.  In this paradigm,
the respective velocity fields of two clouds with extremely different star 
formation efficiencies should 
exhibit very different properties, excluding the effects of feedback. 
The Rosette molecular cloud (RMC) and G216-2.5 molecular cloud provide such 
an extreme pair to evaluate descriptions of star formation. 
The RMC complex
is associated with the 
OB cluster NGC~2244 whose massive star constituents are responsible for the 
excitation of the Rosette Nebula (NGC~2237/2246).   
The Rosette nebula is clearly interacting with the ambient 
cloud and may be responsible for triggering the formation of some of the 
active sites of star formation within the cloud (Cox, Deharveng, \& Leene 1991; 
Williams, Blitz, \& Stark 1995).  A fraction of these 
IRAS Point Sources are comprised of clusters of newborn stars 
(Phelps \& Lada 1997).
The giant molecular cloud G216-2.5 (Maddalena's Cloud) provides a stunning 
contrast to the Rosette Molecular Cloud.   It is larger in size and 
more massive than the Rosette cloud yet it is currently devoid of massive star formation
activity (Maddalena \& Thaddeus 1985; Williams \& Maddalena 1996).

The star formation rates and efficiencies of the two clouds are quantitatively 
distinguished by the respective values of the 
far infrared  (FIR) luminosity to molecular gas mass ratio that is often used as a 
surrogate measure of star formation efficiency as interstellar dust grains, 
heated by the UV radiation of nearby OB stars, emit thermal radiation into the 
FIR bands. For GMCs in the inner Galaxy and Solar neighborhood, 
$L_{FIR}/M_{H_2} \sim 1$ (Scoville \& Good 1989; Mooney \& Solomon 1988). 
The $L_{FIR}/M_{H_2}$ for the 
Rosette and G216-2.5 clouds are 10 and $<$0.07 $L_\odot/M_\odot$ respectively
(Lee, Snell, \& Dickman 1996). 
The lower value is similar to low mass star forming regions in which the 
emitting dust grains are heated exclusively by the interstellar radiation field rather 
than newborn stars (Snell, Heyer, \& Schloerb 1988).
Williams \& McKee (1997) demonstrated that the massive star formation rate
per unit mass in the RMC is an order of magnitude higher than the 
Galactic average and that G216-2.5 is quite unusual in that 90\% of 
clouds of comparable mass are expected to contain at least one O star.

In this paper, we present new \co\ and \coa\ observations of the 
Rosette Molecular Cloud and 
limited imaging of the \co\ and \coa\ J=1-0 
emission from the G216-2.5 molecular cloud 
obtained 
with the Five College 
Radio Astronomy Observatory 14m telescope and SEQUOIA focal plane array 
system. 
These new wide field imaging data provide kinematic information to 
evaluate 
the predictions of the turbulent fragmentation 
and its role 
in regulating the formation of stars both between and within giant molecular 
clouds in the Galaxy. 

\section{Observations}
All observations analyzed in this paper were taken with the 
14m telescope of the Five College Radio Astronomy Observatory.
The FWHM beam size of the antenna 
at the observed frequencies are 45\arcsec\ (115 GHz) and 47\arcsec\
(110 GHz).   
The main beam efficiencies  at these frequencies
are 0.45 and 0.48 respectively as gauged from measurements of
Jupiter.  All intensities reported in this paper are main beam 
temperatures with these efficiencies included (Gordon, Baars, \& Cocke 1992).

The \co\ and \coa\ J=1-0 
observations were taken with the 32 pixel focal plane array 
SEQUOIA using On-the-Fly mapping
in which the telescope is continuously scanned across the source 
while rapidly 
reading the spectrometers.   The data were resampled onto a regular 
20\arcsec\ spaced grid.
Owing to the broad bandwidth of the HEMT 
amplifiers, the \co\ and \coa\ lines were observed simultaneously enabling 
excellent positional registration and calibration.  The backends 
were comprised of a system of 64 autocorrelation spectrometers
each with 50 MHz bandwidth for the Rosette and 25 MHz for G216-2.5.  
No smoothing was applied to the autocorrelation function so the 
spectral resolution was 59 kHz and 29.5 kHz per channel for the Rosette
and G216-2.5 cloud respectively. 
System temperatures ranged from 350-500 K
at the line frequency of the \co\ line (115.271202 GHz) and 150-250 K
at the \coa\ line (110.201353 GHz).  
The median rms values achieved 
in the \co\ and \coa\ data cubes for both clouds are
1.0 K and 0.35 K respectively. 

\section{Results}
\subsection{The Rosette Molecular Cloud}
\begin{figure}
\epsscale{0.75}\plotone{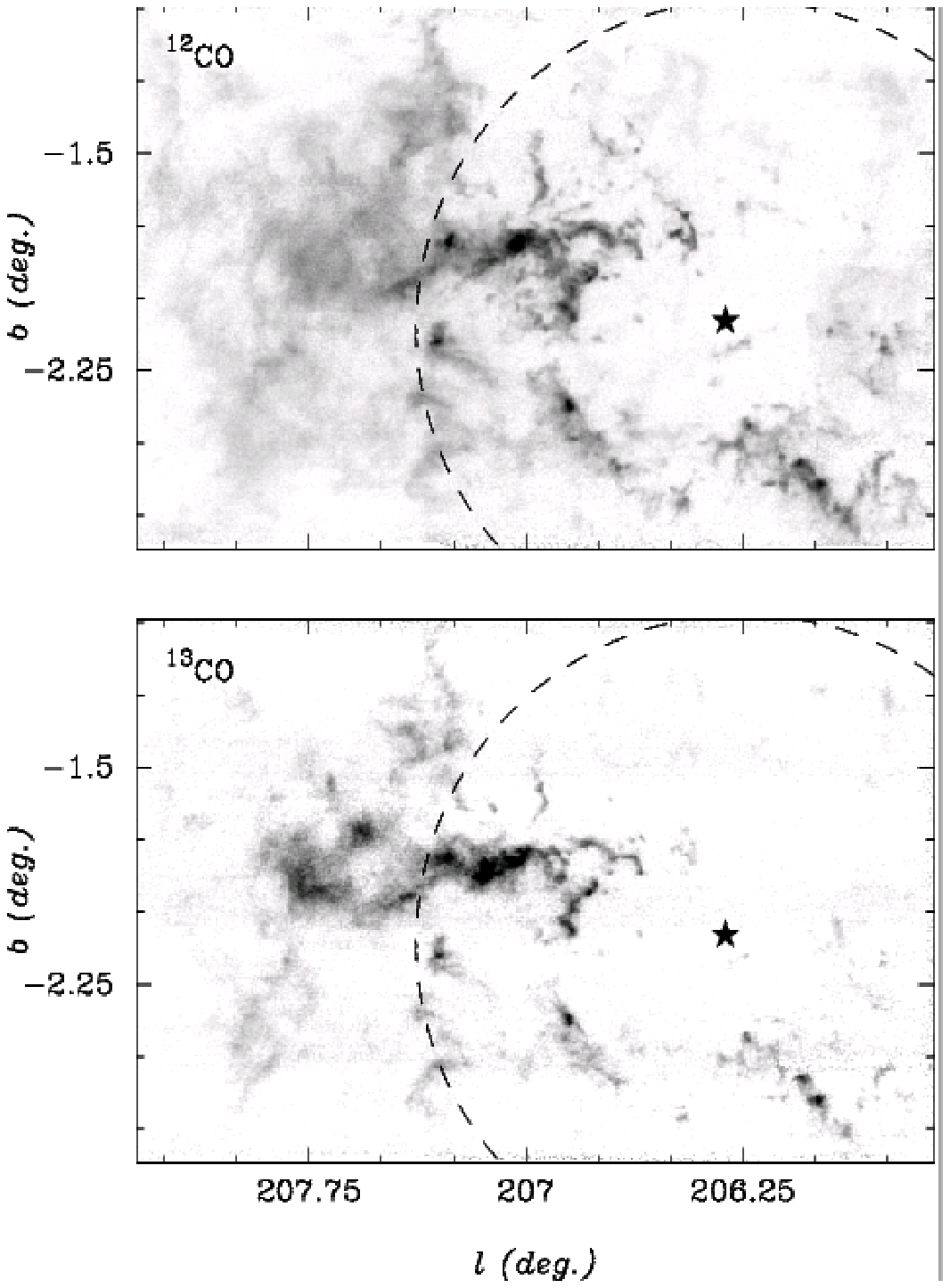}
\caption{Images of the Rosette Molecular Cloud: (top) 
integrated \co\ J=1-0 emission over the velocity 
interval 5 to 25 km s$^{-1}$ (halftone 
ranges from 0 (white) to 100 K kms $^{-1}$ (black); 
(bottom) integrated \coa\ J=1-0 emission 
(halftone 
ranges from 0 (white) to 22 K kms $^{-1}$ (black).    
The dotted line shows the extent of the ionized nebula measured by 
Celnik (1983).   The star denotes the center of the NGC 2244 cluster.
}
\label{figure1}
\end{figure}

The large scale distribution and kinematics of molecular clouds 
are best traced by observations of the low rotational transitions of the 
isotopologues, 
\co\ and \coa.
While the high opacity of \co\ emission precludes a direct 
estimate of molecular hydrogen column density, it enables the line to 
be detected within low column density regimes. \coa\ offers a lower 
opacity tracer of \htwo\ column density but also likely saturates within 
the high density cores.   
Images of integrated \co\ and \coa\ J=1-0 intensities are displayed in 
Figure~\ref{figure1}.  These show the positional relationship between 
the ionization front of the Rosette nebula (dotted line) and the ambient 
molecular material.  
Along the projected boundary of the HII regions, there are both bright, compact
features in both \co\ and \coa\ images that are embedded within a more diffuse component 
seen primarily in \co. 
On many lines of sight towards the nebula, there are patches of visual 
obscuration that are congruent with CO emission features.  These spatial 
coincidences geometrically locate at least part of the cloud along the 
foreground edge of the HII region.   Several of these features are bright 
rimmed
globules that correspond to luminous star forming regions 
within the cloud (Patel \etal 1993; White \etal 1997).

There are textural variations of the \co\ emission across the observed field
and with respect to the projected edge of the ionization front.  Within the 
projected radius of the Rosette nebula, there is
bright, high contrast \co\ emission. 
Exterior to this projected edge, the integrated \co\ emission 
is smoothly distributed and generally quite 
weak ($\int T(^{12}CO)dv$=10-30 K km$s^{-1}$).
The line profiles within the diffuse component
typically show several velocity components separated 
up to 6 \kms\ although the one dimensional velocity dispersion in any 
one component is $\sim$ 1 \kms.   Such composite line shapes are similar to 
those measured in the translucent cloud population that also exhibit diffuse,
low surface brightness distributions of CO emission 
(Magnani, Blitz, \& Mundy 1985).  

Such an extended \co\ component can arise 
from low density, subthermally excited 
molecular material or small scale, high density, structures with a low
beam-filling factor.
The \co\ and \coa\ J=1-0 data presented here can not distinguish which of 
these disparate gas configurations applies.
This diffuse component is less 
prominent within the \coa\ image owing to lower opacities that reduce the 
degree of radiative trapping to excite the line.
Diffuse emission is also present at the low longitude edge of the surveyed area
that is similar to the smooth component external to the nebula. 
The spectra from these regions exhibit much narrower line widths 
than the bright, high contrast 
emission but with velocity centroids consistent with the overall 
rotation of the complex as described by Williams, Blitz, \& 
Stark (1995).  This diffuse component is likely foreground or 
background material that has not yet been perturbed by the 
expanding HII region.

The \co\ and \coa\ intensities are used to estimate the mass and column 
densities of the dominant constituent of molecular hydrogen.
The \co\ luminosity of the cloud is 3.8$\times$10$^4$ $K kms^{-1} pc^2$ that 
corresponds to a molecular mass (including He) of 1.6$\times$10$^5$ \msun\ 
assuming a CO to \htwo conversion of 1.9$\times$10$^{20}$
\htwo\ molecules cm$^{-2}$ (K km s$^{-1}$)$^{-1}$ (Strong \& Mattox 1996).
The diffuse component external to the HII region contributes 46\% of the 
CO luminosity.
\coa\ 
column densities are derived assuming 
local thermodynamic 
equilibrium (LTE) and excitation temperatures derived from 
optically thick \co\ J=1-0 emission 
(Dickman 1978). 
For an \htwo\ to \coa\ abundance of 8$\times$10$^5$, the total mass of the 
mapped area is 1.16$\times$10$^5$ \msun.   
The diffuse component exterior to the HII region has LTE column densities 
of 0.2-1$\times$10$^{22}$ \cmsq.
However, local thermodynamic equilibrium is a 
poor approximation to the excitation of \coa\ line in the diffuse 
component of the Rosette cloud as the population of the upper rotational 
energy levels 
are not properly estimated.  This error underestimates  
the true column densities in this region. 

\subsection{G216-2.5}
\begin{figure}
\plotone{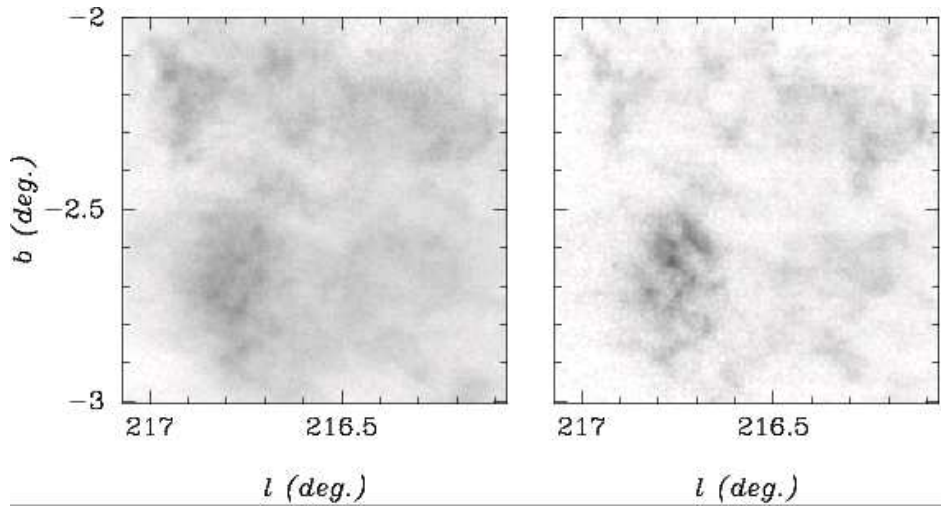}
\caption{Images of (left) \co\ J=1-0 emission  and (right) \coa\ J=1-0 
emission  from the central region of the 
G216-2.5 cloud integrated over the \vlsr\ interval 20 to 30 \kms.  The halftone 
stretch is identical to the \co\ and \coa\ Rosette images in Figure~\ref{figure1}
to demonstrate the low surface brightness emission in G216-2.5.}
\label{figure2}
\end{figure}

The G216-2.5 cloud subtends $>$10 deg$^2$ on the sky (Maddalena \& Thaddeus 1985; Lee, Snell, \& Dickman 1996).  
Owing to time constraints, our new observations cover 1 deg$^2$ centered 
on the brightest \co\ emission identified in these previous 
surveys.
Images of velocity 
integrated emissions are shown in Figure~\ref{figure2}.  The halftone 
range spans the same values as Figure~\ref{figure1} to facilitate a 
direct comparison with the Rosette \co\ and \coa\ images. 
These emphasize the low surface brightness distribution in G216-2.5 
that has been noted by earlier studies (Williams \& Blitz 1998).  
The images show flocculent \co\ emission. 
There are well resolved, coherent clumps that are 
embedded within a smoothly distributed envelope that extends well beyond the 
mapped boundary. The \coa\ emission identifies localized peaks associated with 
the aforementioned \co\ clumps.  
None of these are as bright as the 
core regions within the Rosette cloud.  The molecular line emission from G215-2.5 
is similar to the diffuse component in the Rosette cloud 
that is external to the ionization front. 
As noted by previous studies, there are several 
well resolved velocity components that merge to produce broad line widths 
(FWHM 8.5 km/s).   
\htwo\ column densities are derived from the \co\ and \coa\ data 
assuming LTE and a \htwo\ to \coa\ abundance of 8$\times$10$^5$.  
The mass of molecular hydrogen inclusive of He within the mapped area
is 7$\times$10$^4$ \msun.  

\section{Statistical Measures of Gas Dynamics}
The structure function provides 
a concise, statistical description of 
velocity fluctuations within a fluid volume.  It measures moments of 
velocity differences, ${\delta}v$, as a function of spatial displacement
$\tau$, within a volume. 
The moments of the structure function can be 
recaste as a linear expression and parameterized by a power law 
$$ {\delta}v=v_{\circ}\tau^\gamma \;\;\; \eqno(1) $$
where 
$v_\circ$ is the scaling coefficient and $\gamma$ is the scaling exponent.
Recovering the structure function parameters, $v_\circ, \gamma$, from spectroscopic imaging 
data presents a challenging task to evaluate 
properties of the turbulent gas flow.
The observations do not directly measure 
3D velocity fields but gather a snapshot-in-time view of intensity
at each spectroscopic channel
that is integrated
along the line of sight and
filtered by the
opacity and the excitation of the molecular or atomic line transition.
This transformation requires 
statistical
analyses that can be related to the structure function of the 3D velocity field.
Such analyses have been developed for both 2-dimensional 
moment representations of the data cube (Scalo 1984; Kleiner \& Dickman 
1985; Miesch \& Bally 1994; Stutzki \etal 1998) and the complete 3-dimensional 
data cube (Lazarian \& Pogasyan 2000; Padoan, Goodman, \& Juvela 2003).

Here, we consider Principal Component Analysis to assess the turbulent 
flows within the Rosette and G216-2.5 molecular clouds. 
Principal Component Analysis is a 
powerful method to exploit the informational content within multi-variate
data sets such as the wide field, spectroscopic imaging presented in this 
study (Heyer \& Schloerb 1997; Brunt \& Heyer 2002).  
Data are reordered into a set of eigenvectors and eigenimages that describe the degree of
spectral and spatial correlation of line profile shapes. 
Correspondingly, the eigenvectors and eigenimages 
can be analyzed to recover the statistics of the cloud velocity field as
described by the structure function (Brunt \& Heyer 
2002; Brunt \etal 2003) and place limits on the driving scale of turbulence 
(Brunt 2003).

PCA is applied to the \co\ and \coa\ data cubes of the 
Rosette and G216-2.5 molecular clouds.  
The eigenvectors and eigenimages
derived from the \co\ observations of the two clouds are shown 
in Figure~\ref{figure3} and Figure~\ref{figure4}.
For each principal component, 
the characteristic velocity 
scale, ${\delta}v$, 
and spatial scale, $l$, are calculated from the autocorrelation 
functions of the eigenvector and eigenimage respectively (Brunt \& Heyer 2002).
This step of the analysis has been modified from previous studies 
by fitting an ellipse 
to the 1/e contour of the 2 dimensional ACF to determine the major axis, $l_x$,
minor axis, $l_y$, and orientation with respect to the cardinal directions.  
Small corrections are applied to $l_x$ and $l_y$ to account for
the finite resolution of the observations (Brunt 1999).  The 
spatial scale, $l$, for a given eigenimage is $l=\sqrt{l_x^2+l_y^2}/2^{1/2}$.  
This modification
provides an improved estimate of the spatial scale when the autocorrelation 
function is strongly anisotropic. 
The set of 
${\delta}v,l$ points define a relationship between the magnitude of 
velocity differences and the spatial scale over which these differences
occur that can be related to the velocity structure function (Brunt \etal 2003).
This relationship is parameterized by a power law
with scaling coefficient, $v_\circ$, and scaling exponent, $\alpha_{PCA}$.
This PCA-derived exponent, $\alpha_{PCA}$, is related to the 
scaling coefficient 
of the structure function, $\gamma$, defined in equation (1), 
by an empirically defined relationship 
determined from model velocity fields (Brunt \etal 2003).
$$ \gamma = 1.69(\alpha_{PCA}-0.32) \;\;\;\;for\;\;\; \alpha_{PCA} \le 0.67 \eqno(2)$$
and 
$$ \gamma = 0.93(\alpha_{PCA}-0.03) \;\;\;\;for\;\;\; \alpha_{PCA} > 0.67  \eqno(3)$$
The results for each cloud and CO isotope
are shown in Figure~\ref{figure5} and summarized 
in Table~1.  The power law parameters are derived from a bisector fit to the 
set of ${\delta}v,l$ points.  The parameters uncertainties are estimated 
using the bootstrap Monte Carlo method to estimate the 
underlying probability distribution (Press \etal 1992).  
For both clouds, there is 
no significant difference between the fitted parameters derived
from \co\ and \coa\ data implying that line opacity does not 
play a significant role in the PCA results.  
The lower opacity \coa\ emission 
surely probes deeper into higher column density regions than \co.  
However, such regions do not subtend much solid angle and therefore,
make a small contribution to the overall molecular line luminosity.
Effectively, \co\ and \coa\ recover the same velocity field over the projected 
surface of the cloud (Heyer \& Schloerb 1997; 
Brunt \& Heyer 2002).  

The power law parameters derived for the Rosette and G216-2.5 clouds 
are within the range of values found by 
Heyer \& Brunt (2004) for a set of 27 GMCs
where $<v_\circ>$=0.90$\pm$0.19 
\kms\ and 
$<\gamma>$=0.49$\pm$0.15 that define the universality 
of velocity structure functions within the molecular ISM.  Many of the 
27 molecular clouds are massive reservoirs of material (M$>$10$^5$ \msun) 
and, like the Rosette
cloud, are currently producing 
rich stellar clusters.  The near identical form of individual structure
functions naturally accounts for the well known but variously 
interpreted relationship between cloud size and global velocity 
dispersion identified by Larson (1981).  The similarity of the Rosette and 
G216-2.5 velocity 
structure functions is remarkable given the respective dynamical and evolutionary 
states of the two clouds.  The Rosette cloud is actively forming massive stars and clusters
from which winds and UV radiation have clearly impacted the large scale structure of the cloud.
The G216-2.5 cloud is not self-gravitating and not actively forming early type 
stars, which
precludes a source of internal energy input (Lee, Snell, \& Dickman 1996). 
Yet, 
the derived scaling coefficients, $v_\circ$, are comparable that implicates a
more significant source of energy present at larger scales 
that sustains the turbulent flows within these clouds.  
Energy feedback from protostellar 
winds and HII regions must play a minimal role in the {\it global} dynamics of molecular clouds.

\subsection{Variations of turbulent flow properties within the Rosette Cloud}
\begin{figure}
\plotone{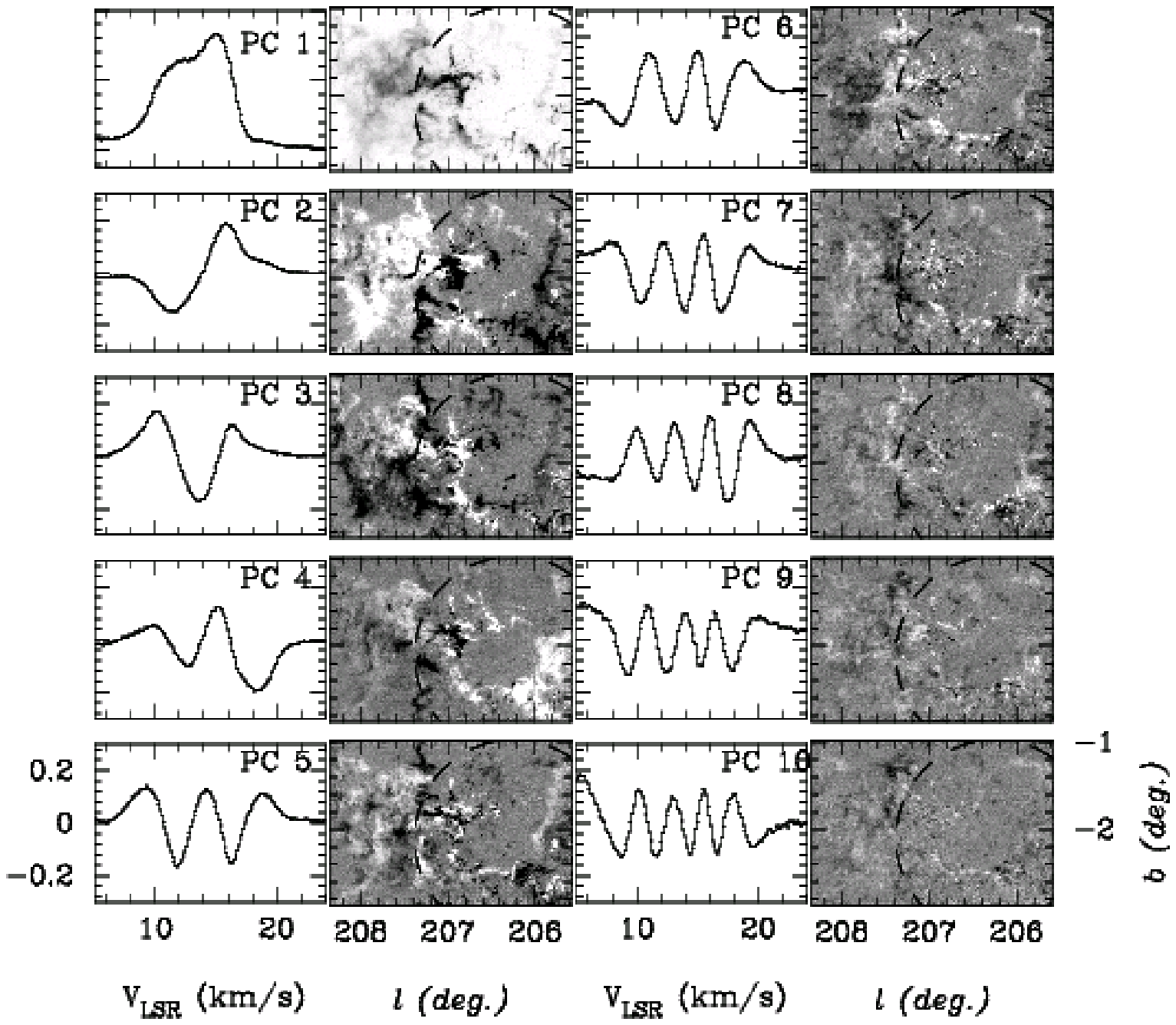}
\caption{The set of eigenvectors and eigenimages derived from the \co\ J=1-0 data cube
of the Rosette Molecular Cloud}
\label{figure3}
\end{figure}

\begin{figure}
\plotone{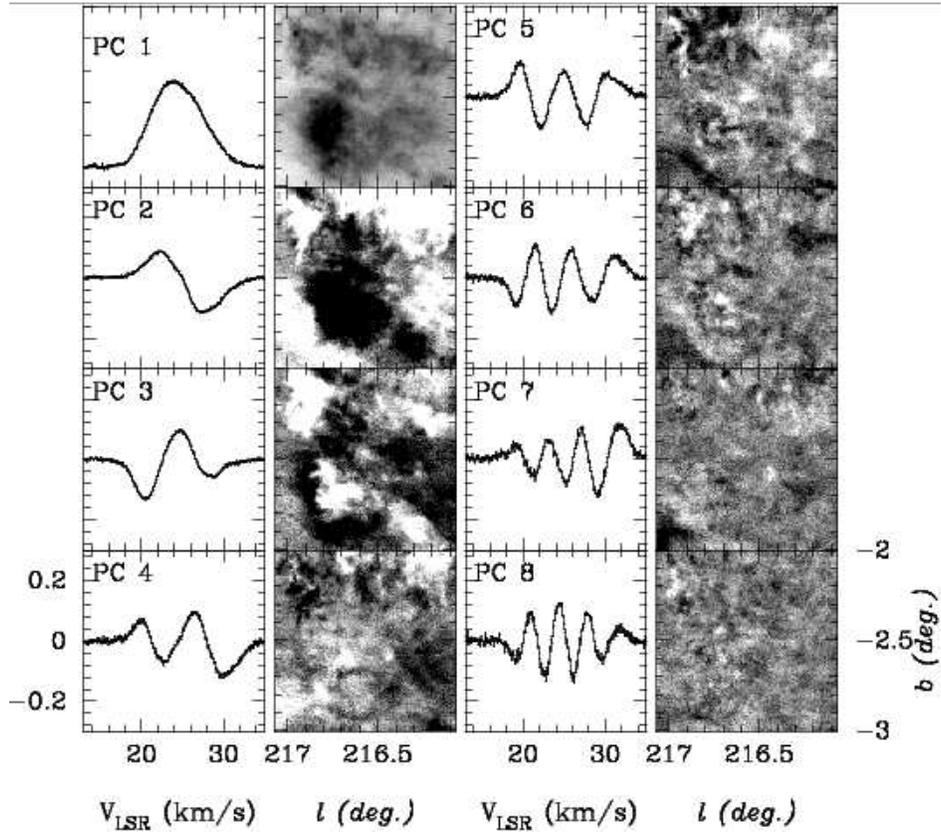}
\caption{The set of eigenvectors and eigenimages derived from the \co\ J=1-0 data cube
of the G216-2.50 molecular cloud}
\label{figure4}
\end{figure}

\begin{figure}
\plotone{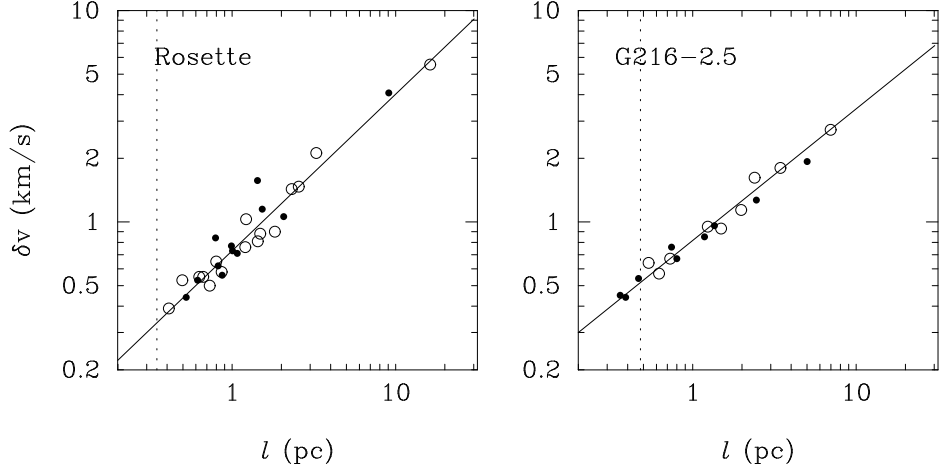}
\caption{The ${\delta}v,l$ relationships for the (left) Rosette molecular 
cloud and (right) G216-2.5 cloud. 
The solid and open circles are 
${\delta}v,l$ pairs from the \co\ and \coa\ data respectively.  
The solid line is the power law fit 
to the \co\ points exclusively.  
The vertical dotted lines shows the spatial 
resolution limit of the observations.
}
\label{figure5}
\end{figure}

\begin{figure}
\plotone{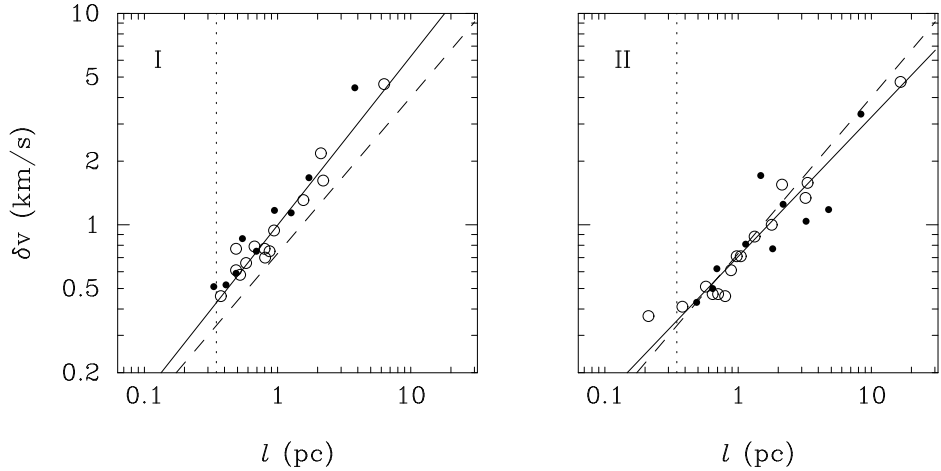}
\caption{The ${\delta}v,l$ relationships derived from \co\ (solid)
and \coa\ (open) data for lines of sight within the projected boundary of
the Rosette ionization front (Zone I) and exterior to the front (Zone II).
The solid line is the power law fit 
to the \co\ points exclusively.  For reference, the dashed line shows the power law fit
for the full cloud shown in Figure~\ref{figure5}.  The structure 
function of molecular gas within the HII region boundary is 
described by a larger 
scaling coefficient and steeper scaling exponent than is found in the 
diffuse extended component.
}
\label{figure6}
\end{figure}

The similarity of velocity structure functions derived over the entire surface of the 
cloud does not preclude local variations of turbulence within a GMC. 
The ionization front of the Rosette Nebula has had a significant 
impact upon one part of the parent molecular cloud 
as manifest by the gas distribution.
In addition, the Rosette \co\ eigenimages in Figure~\ref{figure3} 
reveal varying gas 
dynamics within the cloud that are separated by the ionization 
front.  
For components 4 through 10, 
the eigenimages show high frequency 
power within the ionization front that is distinguished from the 
diffuse, coherent structure  for lines of sight exterior to the front.
We have partitioned the Rosette cloud into two zones that are 
conveniently defined by the 
projected boundary of the HII region to gauge whether the observed textural 
differences, identified in Figure~\ref{figure1} and within the eigenimages 
(Figure~\ref{figure3}), are also reflected in the turbulent 
state of the gas.  
Zone I corresponds to positions within the 
projected radius of the Rosette nebula and Zone II covers the 
area external to the ionization front,  
exclusive of the small area segment in the upper right corner of the 
observed field.   The ${\delta}v,l$ relationships derived from 
the \co\ and \coa\ data for both partitions are shown in Figure~\ref{figure6}
and the 
derived scaling law parameters for both \co\ and \coa\ 
are listed in Table~2. While neither set of parameters deviate beyond the 
range of values found by Heyer \& Brunt (2004), there are significant differences between the two regions.
The Zone I scaling coefficient, $v_\circ$,  
is significantly larger than the corresponding value calculated for Zone II.
The enhanced value
parameters in Zone I may result from the
interaction
between the HII region and cloud.  Such interactions 
inject energy into the system that lead to stronger, non-gaussian 
velocity fluctuations.

The 
scaling law parameters in Zone II are similar to the values derived
for G216-2.5 that indicate a connection between the diffuse gas distribution 
in these regions 
and the properties of the velocity field.  These similarities may be due to 
both regions having recently condensed from overlying atomic material and have 
not yet achieved thermal or dynamical balance.
The absence of resolved high volume 
or column density cores may arise from local velocity fields with 
negligible longitudinal components that are responsible for converging flows.
However, the scaling exponents are 
still larger than the value expected from a purely incompressible flow 
($\gamma$=1/3).
Such smoothly distributed material may also result from 
continued magnetic support enabled by sufficient ionization by ambient 
UV radiation in these low column density zones (Johnstone, Di Francesco,
\& Kirk 2004).  

\section{Discussion}
The decomposition of the Rosette and G216-2.5 spectral line data cubes of \co\ and \coa\ 
emission with PCA offer important constraints to the turbulent flow properties in these 
clouds and enable a coarse but valuable assessment of the turbulent 
fragmentation description of 
star formation.  Specifically, we seek to evaluate the relationship 
of the star formation efficiency with the turbulent flow properties: 
the driving scale, $\lambda_D$, the sonic scale, $\lambda_S$, and the turbulent Mach number that may be expected to regulate the formation and evolution of 
protostellar cores within molecular clouds.
The star formation rate, $dM_*/dt$,
within a given solid angle subtended by the mapped area
is derived from the measured far infrared luminosity,
$$  { {dM_*} \over {dt} } = 
6.3{\times}10^{-10} L_{FIR} \;\;\;\; M_\odot yr^{-1} \eqno(4) $$
where
$$
L_{FIR}=4{\pi}d^2\Sigma {\Delta}\nu_i S_{\nu_i}  \;\;\; \eqno(5)
$$
${\Delta}\nu_i$ and $S_{\nu_i}$ are the bandwidth and measured 
IRAS fluxes at 60 and 100$\mu$m 
bands, d is the distance to the cloud
and a Miller-Scalo IMF is assumed
(Thronson \& Telesco 1986).
The fluxes include contributions from both embedded point sources 
and extended emission from grains heated by strong, UV fields of
massive stars.  Consequently, it provides an {\it upper limit} to
the star formation rate for
low mass star forming regions where the diffuse far-infrared dust component is
primarily heated by the ambient
interstellar
radiation field.  
The inverse gas depletion time, $t_g^{-1}$, is derived from the ratio of the cloud mass to 
the star formation rate.  Its inverse, $t_g^{-1}$, 
provides a surrogate measure 
of the star formation efficiency. 
Using \co\ to derive M(\htwo), the values for $t_g^{-1}$ are 
2.9$\times$10$^{-9}$ yr$^{-1}$ for the Rosette and $<$1.8$\times$10$^{-10}$ 
yr$^{-1}$ for G216-2.5.

The decomposition of the \co\ and \coa\ data cubes with PCA are used to 
derive the basic turbulent flow properties to compare with the star formation efficiencies.  
The
role of the turbulent driving scale is not considered here
as there is no accurate measure of $\lambda_D$
but only coarse
estimates that $\lambda_D$ is comparable to or larger than
the size of the cloud
(Brunt 2003).  

The sonic scale, $\lambda_S$, corresponds to the scale at which the
turbulent velocity fluctuations are comparable to the local sound speed.
It can be observationally defined from 
the scaling parameters, $v_\circ$ and $\gamma$, 
$$ {\delta}v(\lambda_S) = v_\circ{\lambda_S}^\gamma = c_s \eqno(5) $$
where $c_s$ is the local 1D, isothermal sound speed for \htwo\ inclusive
of He abundance.  Inverting this expression, the sonic scale is,
$$ \lambda_S = (kT/v_\circ^2{\mu}m_{H_2})^{1/2\gamma} \eqno(6) $$
where T is the kinetic temperature of the gas and $\mu$ is the mean 
molecular weight.  
Values of $\lambda_S$ for the Rosette 
and G216-2.5 are derived for each isotope 
assuming a 
kinetic temperature of 10 K and are listed in Table~1.
This may underestimate the sonic scale for the Rosette for which 
warmer temperatures 
may prevail within the star or cluster forming cores (Williams \& Blitz 1998).
However, such cores comprise a small fraction of the total mass 
of the cloud.  The \co\ peak temperatures for much of the Rosette 
cloud are consistent with 10 K.   
Based on the \co\ data, the sonic scales  for the 
Rosette and G216-2.5 clouds are 
0.33$\pm$0.02 pc and  0.20$\pm$0.04 pc respectively.

The turbulent Mach number is conventionally 
estimated from the dispersion of velocities within a cloud volume.  
Formally, this global line width 
simply reflects the structure function evaluated at the cloud size and is not 
useful when comparing clouds with different sizes. 
For example, two clouds with identical structure 
functions but with different sizes, would have different global 
velocity dispersions.
Numerical simulations of interstellar turbulence often 
quote the variance of the full
velocity field.   However, such studies typically keep 
the size of the model cloud 
constant so these effectively evaluate the Mach number at a fixed scale.
A more useful quantity to compare clouds is the mean 
turbulent velocity fluctuation at a fixed scale.  The scaling coefficient, $v_\circ$,
defined in equation 1, describes the amplitude of velocity fluctuations at a scale 
of 1 pc.  We use this quantity to derive the Mach number, $M_{1pc}$ measured at 
this fixed reference scale,
$$ M_{1pc} = v_\circ/\sqrt{kT/{\mu}m_{H_2}}  \eqno(7) $$
The $M_{1pc}$ is 4.2$\pm$0.17 for the Rosette and 4.7$\pm$0.12 for G216-2.5 
assuming a kinetic temperature of 10 K.

\begin{figure}
\plotone{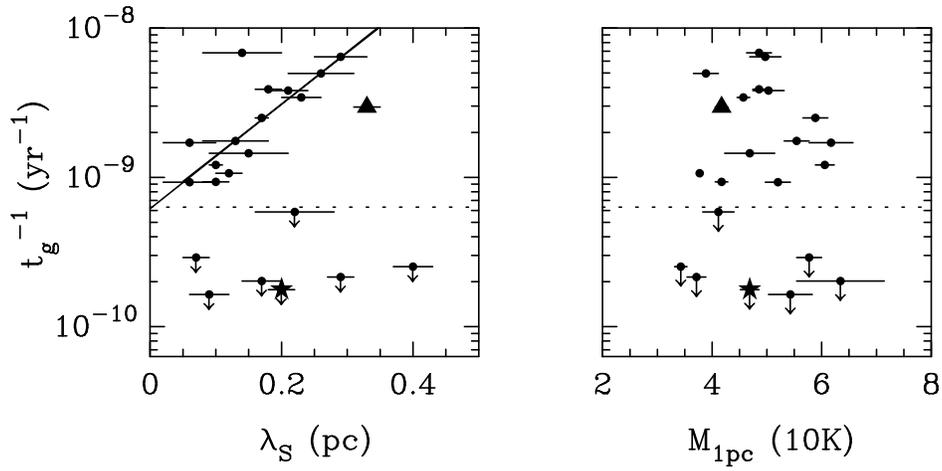}
\caption{A semi-log plot of the inverse gas depletion time, t$_g^{-1}$, 
with
(left) the sonic scale, $\lambda_S$ and (right) Mach number measured at 1 pc scale with 
a kinetic temperature of 10 K for a set of giant molecular clouds.  The Rosette
cloud is shown as a filled triangle and the G216-2.5 cloud is 
plotted as a filled star.
No significant correlation exists 
between the star formation activity and turbulent flow properties for the 
full set of clouds contrary to 
predictions by 
turbulent fragmentation descriptions.  
A trend is identified (solid line) 
between t$_g^{-1}$ and $\lambda_S$ for the 
subset of clouds with 
L$_{FIR}$/M(H$_2$) $>$ 1.
}
\label{figure7}
\end{figure}

From this limited sample of two extremely different clouds, 
one could 
conclude that 
the turbulent Mach number does not regulate the formation of stars in 
molecular clouds and there is some positive relationship 
between star formation efficiency and $\lambda_S$.  
This assumes that 
our measure of the star formation efficiency, $t_g^{-1}$, 
is indicative of 
the fraction of the cloud that is converted into stellar mass after several 
crossing times. 
To more critically 
examine this relationship, 
we have 
supplemented the Rosette and G216-2.5 clouds with
\co\ measurements from GMCs within 
the FCRAO Outer Galaxy Survey (Heyer \etal 1998) and recent imaging surveys
of the Cepheus region of the Galaxy (Brunt \& Mac Low 2004).  These additional clouds 
span the range of star formation efficiencies set by the Rosette and G216-2.5 
clouds and comprise the sample used by Heyer \& Brunt (2004).  
For GMCs with $L_{FIR}/M(H_2) < 1$, we designate the derived values of the 
inverse gas depletion time to be upper limits 
as a significant fraction of the emitted far infrared flux is 
from grains heated by the ambient radiation field rather than 
nearby, recently formed early type stars.
The variation of $t_g^{-1}$ with 
$\lambda_S$ and $M_{1pc}$ is shown 
in Figure~\ref{figure7}.   For the full set of clouds, no systematic 
trends can be identified between the 
star formation efficiency and either the sonic scale or Mach number.

A relationship between $t_g^{-1}$ 
and $\lambda_S$ can be 
identified within the set of clouds with $L_{FIR}/M(H_2) > 1$ that 
are currently producing OB stars and stellar clusters.  The relationship is 
described by the fitted exponential,
$t_g^{-1}=(6.3\pm1.9){\times}10^{-10}exp((8.0\pm1.7){\lambda}_S)$ yr$^{-1}$
that demonstrates a 
sensitive dependence of the star formation efficiency 
on the 
sonic scale for these regions.   However, 
the ratio of newborn stellar mass to that of the parent molecular cloud  for
this sample of clouds is 
low (several percent). 
Such low star formation efficiencies 
are realized in numerical simulations of turbulence 
for which the driving scale 
is small compared to the cloud size enabling wide-spread 
support of the 
cloud against self-gravity (Klessen, Heitsch, \& Mac Low 2000).  
In the non-magnetic case, simulations that drive the turbulent 
motions at large scales are overly efficient 
at producing newborn stars (Klessen, Heitsch, \& Mac Low 2000; 
Vazquez-Semadeni, 
Ballesteros-Paredes, \& Klessen 2003). 
Li \etal (2004) show that MHD
turbulence driven at large scales can delay the onset collapse and 
reduce the efficiency.  
Similarly, Vazquez-Semadeni \etal (2005) 
demonstate that the efficiencies are similar to observed 
values by the inclusion of the 
magnetic field that limits the probability of forming 
collapsing cores.   

These results dictate that non-magnetic,  
turbulent fragmentation does not exclusively 
regulate the formation of stars in the ISM.   It may play a role in 
determining the efficiency through which stars are produced within 
molecular clouds currently forming clusters and massive stars. 
Such regions 
may be globally supercritical, 
($(M/\Phi)/(M/\Phi)_{crit}>>1$) 
so that density perturbations can readily develop 
from supersonic 
shocks generated by turbulent gas flows or from external triggering agents
such 
as an expanding ionization front.   For low mass star forming regions,
the role of non-magnetic turbulent fragmentation must necessarily be limited 
in this sample 
of clouds. 

\section{Summary}
Wide field imaging of \co\ and \coa\ J=1-0 emission from the Rosette and 
G216-2.5 molecular clouds is presented.  These GMCs represent quite different states of 
star formation activity and provide valuable information to assess interstellar 
turbulence and its role in regulating star formation. 
From the analysis of the spectroscopic data, we find
\begin{enumerate}
\item The velocity fields of both clouds are described by 
structure functions with similar scaling parameters.  
This suggests that 
turbulent motions are sustained by an external source of energy and 
not by internal energy sources.
\item Within the Rosette Molecular cloud, we identify local variations 
of the velocity structure function due to the expansion of ionized gas.
However, such interactions are spatially limited and have not 
yet propagated through the cloud to significantly modify the 
global dynamics of the cloud.  
\item None of the relationships predicted by turbulent fragmentation 
descriptions of star formation are identified within the full set of giant 
molecular clouds that span a broad range of masses and specfic star formation 
rates, 
$L_{FIR}/M(H_2)$.   For a subset of clouds that are currently 
forming clusters, a correlation is found between the sonic scale and
the inverse gas depletion time, $t_g^{-1}$.  We conclude that 
turbulent fragmentation 
does not exclusively regulate star formation 
activity within all clouds. 
\end{enumerate}

\acknowledgments
This work was supported by NSF grant AST 02-28993  to the Five College
Radio Astronomy Observatory.    JPW acknowledges support from NSF grant 
AST 03-24328. 
CB holds an RCUK Academic Fellowship at the University of Exeter.

\clearpage

\begin{table}[htb]
\label{table1}
\begin{center}
\caption{Velocity Structure Function Parameters }
\vspace{7mm}
\begin{tabular}
{lccccc}
\hline
 & $v_\circ$  & $\alpha_{PCA}$ & $\gamma_1$ & $\lambda_S$(10 K)\\
 &            &                &            & (pc)\\
\hline
Rosette \\
\hspace {.25in} \co\ & 0.73 $\pm$ 0.03 & 0.74 $\pm$ 0.04 & 0.66$\pm$0.04 & 0.33$\pm$0.02 \\
\hspace {.25in} \coa\ & 0.77 $\pm$ 0.04 & 0.84 $\pm$ 0.13& 0.76$\pm$0.12 & 0.36$\pm$0.07\\
G216-2.5\\
\hspace {.25in} \co\ & 0.82 $\pm$ 0.03 & 0.63 $\pm$ 0.04 & 0.52$\pm$0.07 & 0.20$\pm$0.04 \\
\hspace {.25in} \coa\ & 0.80 $\pm$ 0.02 & 0.56 $\pm$ 0.02 & 0.40$\pm$0.04 & 0.13$\pm$0.02 \\
\hline
\end{tabular}
\end{center}
\end{table}

\begin{table}[htb]
\label{table2}
\begin{center}
\caption{Velocity Structure Function Parameters within Rosette Cloud}
\vspace{7mm}
\begin{tabular}
{lcccc}
\hline
 & $v_\circ$  & $\alpha_{PCA}$ & $\gamma_1$ \\
\hline
Rosette Zone I \\
\hspace {.25in} \co\ & 1.00 $\pm$ 0.04 & 0.79 $\pm$ 0.06 & 0.71$\pm$0.05 \\
\hspace {.25in} \coa\ & 1.16 $\pm$ 0.07 & 0.86 $\pm$ 0.09 & 0.78$\pm$0.08 \\
Rosette Zone II \\
\hspace {.25in} \co\ & 0.70 $\pm$ 0.03 & 0.66 $\pm$ 0.06 & 0.58$\pm$0.10 \\
\hspace {.25in} \coa\ & 0.68 $\pm$ 0.07 & 0.67 $\pm$ 0.12 & 0.59$\pm$0.20 \\
\hline
\end{tabular}
\end{center}
\end{table}

\end{document}